\documentclass[fleqn,twoside]{article}
\usepackage{espcrc2}
\usepackage{epsfig}

\newcommand\as{\alpha_s}
\newcommand\f[2]{\frac{#1}{#2}}

\def\nn{\nonumber}

\def\b0{b_0}

\title{QCD Corrections to Dilepton Production
near Partonic Threshold in $\bar{p}p$ 
Scattering\thanks{Presented by H.\ Yokoya at the 
``7th International Symposium on Radiative Corrections (RADCOR 2005)'', 
October 2-7, 2005, Shonan Village, Japan}
}
\author{
H.\ Shimizu\address{
 Department of Physics, Hiroshima University, 
 Higashi-Hiroshima 739-8526, Japan},
G.\ Sterman\address{
 C.N.\ Yang Institute for Theoretical Physics,
 Stony Brook University,\\
 Stony Brook, New York 11794 -- 3840, U.S.A.},
W.\ Vogelsang\address{
 Physics Department and RIKEN-BNL Research Center, 
 Brookhaven National Laboratory,\\
 Upton, New York 11973, U.S.A.}
and H.\ Yokoya\address{
 Department of Physics, Niigata University, Niigata 950-2181, Japan}
}

\begin{document}

\begin{abstract}
We present a recent study of the QCD corrections to dilepton
 production near partonic threshold in transversely polarized $\bar{p}p$
 scattering. 
We analyze the role of the higher-order perturbative QCD corrections in
 terms of the available fixed-order contributions as well as of
 all-order soft-gluon resummations for the kinematical regime of
 proposed experiments at GSI-FAIR. 
We find that perturbative corrections are large for both unpolarized
 and polarized cross sections, but that the spin asymmetries are stable. 
The role of the far infrared region of the momentum integral in the
 resummed exponent and the effect of the NNLL resummation are briefly
 discussed. 
\end{abstract}

\maketitle

\begin{picture}(5,2)(-370,-425)
\put(0,-95){BNL-NT-06/7}
\put(0,-110){HUPD-0604}
\put(0,-125){RBRC-585}
\put(0,-140){YITP-SB-06-03}
\end{picture}

\vspace*{-20pt}

\section{INTRODUCTION}
A polarized antiproton beam of energy $E_{\bar{p}}=15\!-\!22$~GeV
may be available in future experiments at the GSI-FAIR project. 
Measurements of dilepton production in transversely polarized $\bar{p}p$
collisions are the main motivation for the proposed GSI-PAX~\cite{pax}
and GSI-ASSIA~\cite{assia} experiments. The measurements would be
carried out using a transversely polarized fixed proton target, or a
proton beam of moderate energy $E_{p}=3.5$ GeV. 

Measurements of the transverse double-spin asymmetry 
\begin{equation}
A_{TT}\equiv\frac{\delta{\sigma}}{\sigma}=\frac{\sigma^{\uparrow\uparrow}
 -\sigma^{\uparrow\downarrow}}
 {\sigma^{\uparrow\uparrow}+\sigma^{\uparrow\downarrow}} \; ,  
\end{equation}
 defined as the ratio of transversely polarized and
 unpolarized cross sections, may provide information of the transversely
 polarized parton distribution functions of the proton, dubbed ``transversity''
 $\delta{f}$~\cite{ralston,jaffeji}. Transversity will be probed by
measurements of $A_{TT}$ in polarized $pp$ collisions at the
BNL-RHIC collider~\cite{rhic}. However, since the $\delta{f}$ for sea quarks 
are expected to be small, the asymmetry is estimated to be at most a few
percent~\cite{attnlo}. In contrast, since for the Drell-Yan process 
in $\bar{p}p$ collisions the scattering of two valence quark densities 
contributes, and since in addition the kinematical
regime of the planned GSI experiments is such that rather large parton 
momentum fractions $x\sim 0.5$ are relevant, 
a very large $A_{TT}$ of order 40$\%$ or more is 
expected~\cite{anselmino,shimizu,others}. 
Therefore, unique information on
transversity in the valence region may be obtained from the GSI measurements,
and information from RHIC and the GSI would be complementary.

Here we give a brief report on a recent study of perturbative-QCD 
corrections to the cross sections and to $A_{TT}$ for Drell-Yan dilepton production
at GSI-FAIR~\cite{shimizu}. We discuss the available fixed order corrections as well 
as all-order soft-gluon ``threshold'' resummation.

\section{DRELL-YAN CROSS SECTIONS}
By virtue of the factorization theorem, the cross section for the Drell-Yan
process at large lepton pair invariant mass $M$ can be written in
terms of a convolution of parton distribution functions and
partonic scattering cross sections:
\begin{eqnarray}
\label{fact}
&&\frac{d(\delta)\sigma}{dM^2d\phi}=\sum_{a,b}\int_\tau^1 \!dx_a
 (\delta)\!f_a(x_a,\mu^2)\\ 
 &&\hspace{5pt}\times\int_{\tau/x_a}^1\!\!\!\!dx_b(\delta)\!f_b(x_b,\mu^2)
\frac{d(\delta)\hat{\sigma}_{ab}}{dM^2d\phi} + 
{\cal O}\!\left(\frac{\lambda}{M}\right)^p \, , \nn
\end{eqnarray}
where $\tau=M^2/S$ with $S$ the hadronic c.m.\ energy, and where $\phi$ is
the azimuthal angle of one of the leptons. $\mu$ is the factorization scale. 
As indicated in Eq.~(\ref{fact}),
there are corrections suppressed with some power $p$ and some
hadronic scale $\lambda$. These corrections will become important for
small $M$ and in particular for lower-energy collisions.  
 
\subsection{Fixed-order perturbative calculation}
The partonic cross section is calculated in QCD perturbation theory as a
series in $\alpha_s$;
\begin{eqnarray}
&&\frac{d(\delta)\hat\sigma_{ab}}{dM^2d\phi} = (\delta)\hat{\sigma}^{(0)}_{ab}
\left[\omega_{ab}^{(0)}(z)+\frac{\alpha_s}{\pi}(\delta)\omega^{(1)}_{ab}(z,r)
\right.\nn\\&&\hspace{40pt}\left.
+\left(\frac{\alpha_s}{\pi}\right)^2(\delta)\omega^{(2)}_{ab}(z,r)
+\ldots \right]\ ,
\end{eqnarray}
where $z=M^2/\hat{s}$, $\hat{s}=x_a x_b S$ and $r=M^2/\mu^2$.
For the unpolarized cross section the calculation has been performed up to ${\cal
O}(\alpha_s^{2})$~\cite{vN}, for the transversely polarized case
to ${\cal O} (\alpha_s)$~\cite{wv}. The lowest order gives
\begin{equation}
 \hat\sigma^{(0)}_{q\bar{q}} = \frac{2\alpha^2 e_q^2}{9M^2\hat{s}}\,,\,\,\,\,
\delta\hat\sigma^{(0)}_{q\bar{q}} = \frac{\alpha^2 e_q^2}{9M^2\hat{s}}
\cos{2\phi} 
\end{equation}
with $\omega^{(0)}_{q\bar{q}}=\delta{(1-z)}$.
The higher-order functions may be found in the literature~\cite{vN,wv}. 

\subsection{Threshold resummation}
Threshold resummation addresses large logarithmic perturbative corrections
to the partonic cross section that arise when the initial partons
have just enough energy to produce the lepton pair. Only emission of 
relatively soft gluons is allowed in this case. The large corrections
exponentiate when Mellin moments of the partonic cross section, defined as
\begin{equation}
 (\delta)\omega^{(k),N}_{q\bar{q}}(r) = \int^1_0 dz\,z^{N-1}
  (\delta)\omega^{(k)}_{q\bar{q}}(z,r)\,,
\end{equation}
are taken. To next-to-leading logarithmic (NLL) accuracy one then has for
the resummed cross section~\cite{sterman,catani}:
\begin{eqnarray}
\label{dyres}
&&(\delta) \omega_{q\bar{q}}^{{\rm res},N}(r,\as(\mu)) =
\exp{\left[C_q (r,\as(\mu))\right]}\\
&&\hspace{50pt}\times\exp\left\{ 2 \int_0^1 dz\,
			  \f{z^{N-1}-1}{1-z}\right.\nn\\  
&&\hspace{50pt}\times\left.\int_{\mu^2}^{(1-z)^2 M^2} \f{dk_T^2}{k_T^2}
       A_q(\as(k_T))\right\} \; ,\nn 
\end{eqnarray}
where 
\begin{eqnarray}
\label{andim}
A_q(\as)=\frac{\as}{\pi} A_q^{(1)} +  
\left( \frac{\as}{\pi}\right)^2 A_q^{(2)} + 
\ldots \; ,
\end{eqnarray}
with $A_{q}^{(1)}=C_F$ and~\cite{KT}:
\begin{eqnarray}\label{A12coef}
A_q^{(2)}=\frac{C_F}{2}\left[ 
C_A\left(\frac{67}{18}-\frac{\pi^2}{6}\right)  
-\frac{5}{9}N_f\right]\ ,
\end{eqnarray} 
where $N_f$ is the number of flavors and $C_A=3$. 
The coefficient $C_q (r,\as(\mu))$ collects mostly 
hard virtual corrections. It is given as
\begin{eqnarray}
C_q (r,\as)=\frac{\alpha_s}{\pi}\!
\left(\!-4+\!\frac{2\pi^2}{3}\!+\!\frac{3}{2}\ln{r}\right)\!+{\cal O}(\alpha_s^2) .
\end{eqnarray}
We note that it was shown in~\cite{Eynck:2003fn} that these coefficient
functions also exponentiate. 

Eq.~(\ref{dyres}) is ill-defined because
of the divergence in the perturbative running coupling 
$\alpha_s(k_T)$ at $k_T=\Lambda_{\rm QCD}$. The perturbative 
expansion of the expression shows factorial divergence, 
which in QCD corresponds to a power-like ambiguity of the series.
It turns out, however, that the factorial divergence appears only
at nonleading powers of momentum transfer. The large logarithms
we are resumming arise in the region~\cite{catani} 
$z\leq 1-1/\bar{N}$ in the integrand in Eq.~(\ref{dyres}). Therefore 
to NLL they are contained in the simpler expression
\begin{eqnarray} \label{dyres1}
&&2 \int_{M^2/\bar{N}^2}^{M^2} \f{dk_T^2}{k_T^2} A_q(\as(k_T))
\ln\frac{\bar{N}k_T}{M}
\end{eqnarray}
for the second exponent in~(\ref{dyres}). Here we have chosen
$\mu=M$. This form, to which we will return below, is used for ``minimal'' 
expansions~\cite{Catani:1996yz} of the resummed exponent. 

For the NLL expansion of the resummed exponent one finds 
from Eqs.~(\ref{dyres}),(\ref{dyres1})~\cite{Catani:1996yz}:
\begin{eqnarray}
\label{lndeltams}
&&\ln \delta \omega_{q\bar{q}}^{{\rm res},N} (r,\as(\mu))
= C_q (r,\as(\mu)) \\ 
&&\hspace{55pt}+ 2\ln \bar{N} \;h^{(1)}(\lambda) +
2 h^{(2)}(\lambda,r) \; ,\nn
\end{eqnarray}
where 
\begin{equation}  \label{lamdef}
\lambda=\b0 \as(\mu) \ln \bar{N} \;.
\end{equation}
The explicit expressions for the functions $h^{(1)}$ and $h^{(2)}$ can be
found e.g. in Refs.~\cite{Catani:1996yz,shimizu}.

The hadronic cross section is obtained by performing an
inverse Mellin transformation of the resummed partonic cross section, 
multiplied by the appropriate moments of two parton densities:
\begin{eqnarray}
\frac{d(\delta)\sigma^{\rm res}}{dM^2d\phi}&=&\int_{C}\frac{dN}{2\pi{\it
  i}}\tau^{-N}\nonumber \\
&\times&\sum_{ab} (\delta)f_a^{N}(\delta)
f_b^{N}\frac{d(\delta)\hat\sigma^{{\rm res},N}_{ab}}{dM^2d\phi}\, .
\end{eqnarray}    
In order to perform the inverse Mellin integral, we need to specify a
 prescription for dealing with the singularity in the perturbative
 strong coupling constant in Eq.~(\ref{dyres}). 
We will use the minimal prescription developed in
 Ref.~\cite{Catani:1996yz}, which relies on use of the NLL expanded
 form involving the $h^{i}(\lambda)$, and on choosing a contour to the 
left of the Landau
 singularity at $\lambda=1/2$ in the complex-$N$ plane.
   
\begin{figure}[t]
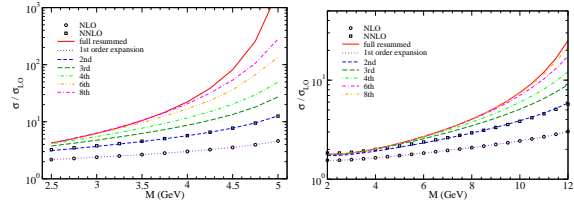

\begin{center}
\epsfig{figure=kfac-fixed.eps,width=.23\textwidth}
\epsfig{figure=kfac-s210.eps,width=.23\textwidth}
\vspace{-40pt}
\caption{$K$-factors as defined in Eqs.~(\ref{eq:kfac}), (\ref{ksoftg}) for
 the Drell-Yan cross section as a function of lepton invariant mass $M$,
 in $\bar{p}p$ collision with $S= 30$ GeV$^{2}$ (left), and $S=$ 210
 GeV$^{2}$ (right). }\label{kfac} 
\end{center}
\vspace{-30pt}
\end{figure}

Figure~\ref{kfac} shows the effects of the higher orders
 generated by resummation for $S=30$~GeV$^2$ and $S=210$~GeV$^2$. 
 We define a resummed ``$K$-factor'' as the ratio of the resummed cross
 section to the leading order (LO) cross section,  
\begin{equation}
\label{eq:kfac}
K^{{\rm (res)}} = \f{{d\sigma^{\rm (res)}}/{dMd\phi}}
{{d\sigma^{\rm (LO)}}/{dMd\phi}}\, ,
\end{equation}
 which is shown by the solid line in Fig.~\ref{kfac}. 
As can be seen, $K^{{\rm (res)}}$ is very large, meaning that
 resummation results in a dramatic enhancement over LO,
 sometimes by over two orders of magnitude for the collisions at lower 
energy. It is then interesting to see how this enhancement builds up
 order by order in perturbation theory. 
We expand the resummed formula to next-to-leading order (NLO) and beyond and define the
 ``soft-gluon $K$-factors'' 
\begin{equation} 
\label{ksoftg}
K^n\;\equiv\; \f{{\left. d\sigma^{\rm{(res)}}/{dMd\phi}\right|_{{\cal O}
(\as^{n})}}}{{d\sigma^{\rm (LO)}}/{dMd\phi}} \; ,
\end{equation}
 which for $n=1,2,\ldots$ give the effects due to 
 the ${\cal O}(\as^{n})$ terms in the resummed formula.
The results for $K^{1-8}$ are also shown in Fig.~\ref{kfac}. 
One can see that there are very large contributions even 
 beyond NNLO, in particular at the higher $M$. Clearly, the
 full resummation given by the solid line
 receives contributions from high orders.
We stress that the ${\cal O}(\alpha_s)$ and ${\cal
 O}(\alpha_s^2)$ expansions of the resummed result
are in excellent agreement with
the full NLO and NNLO ones, respectively 
(circle and square symbols in Figure~\ref{kfac}). This 
shows that the higher-order corrections are really dominated by the
 threshold logarithms and that the resummation is accurately collecting the
latter.
   
\subsection{Far infrared cut-off}
There is good reason to believe that the large enhancement from soft-gluon 
radiation seen above is only partly physical.  
The large corrections arise from a region where the integral in the 
 exponent becomes sensitive to the behavior of the integrand at small 
 values of $k_T$. 
As long as $\Lambda_{\mathrm QCD}\ll M/ \bar{N}\ll M$, the use of
 perturbation theory may be justified, but when $|N|$ becomes very large,
$k_T$ will reach down to nonperturbative scales. 
We seek a modification of the perturbative expression in Eq.~(\ref{dyres})
 that excludes the region in which the absolute value of $k_T$ is less
 than some nonperturbative scale $\mu_0$.
To implement this idea, we will adopt a modified resummed hard
 scattering, which reproduces NLL logarithmic behavior in the moment
 variable $N$ so long as $M/\bar{N}>\mu_0$, but ``freezes'' once 
 $M/\bar{N}<\mu_0$. 
If nothing else, this will test the importance of the region $k_T \le
 \Lambda_{\mathrm  QCD}$ for the resummed cross section. 
If $N$ were real and positive, we could simply replace the resummed
 exponent in~(\ref{dyres1}) by 
\begin{eqnarray}
\label{dyres2}
4 \int_{\rho(M/\bar{N},\, \mu_0)}^M \f{dk_T}{k_T} A_q(\as(k_T))
\ln\frac{\bar{N}k_T}{M} \; ,
\end{eqnarray}
where $\rho(a,b)=\max(a,b)$, and where $\mu_0$ then serves to cut off
the lower logarithmic behavior. 
To provide an expression that can be analytically 
continued to complex $N$, we choose 
$\rho(a,b)=(a^p+b^p)^{1/p}$, with integer $p$. 
This simple form is consistent with the minimal expansion given above,
 and it also allows for a straightforward analysis of the ensuing branch
 cuts in the complex-$N$ plane. For definiteness, we choose $p=2$. 
We will continue to use the expansions in Eq.~(\ref{lndeltams}), but
 redefine $\lambda$ in Eq.~(\ref{lamdef}) by
\begin{eqnarray}
\label{lamdef1}
\lambda=\b0\as(\mu)\ln\bar{N}\!-\!\frac{1}{2}
\b0 \as(\mu)\ln\!\left(\!1\!+\!\frac{\bar{N}^2\mu_0^2}{M^2} 
\right). 
\end{eqnarray}
Of course, different choices of $\mu_0$ give different results, but we should think
 of $\mu_0$ as a kind of factorization scale, separating perturbative
 contributions from nonperturbative. 
Thus changes in $\mu_0$ would be compensated by changes in a
 nonperturbative function. 
Our interest here, however, is simply to illustrate the modification of
 the perturbative sector, which we do by choosing $\mu_0=0.3$ GeV and
 $\mu=0.4$ GeV. 

\begin{figure}[b]
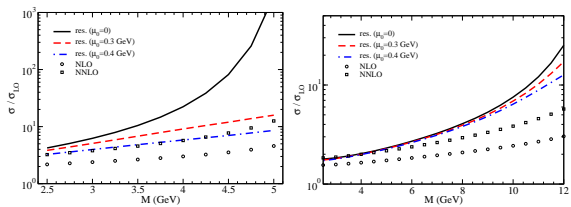

\vspace{-10pt}
\begin{center}
\epsfig{figure=kfac-ps-s30-rev.eps,width=.23\textwidth}
\epsfig{figure=kfac-ps-s210-rev.eps,width=.23\textwidth}
\vspace{-40pt}
\caption{$K$-factors as in Fig.~\ref{kfac}, at $S=30$
 GeV$^{2}$ (left) and $S=210$ GeV$^{2}$ (right). 
The dashed (dot-dashed) lines show the effects of a lower cutoff
 $\mu_0=300$ MeV (400 MeV) for the $k_T$ integral in the
 exponent.}\label{farir}
\end{center}
\vspace{-30pt}
\end{figure}
Results for the ``$K$-factor'' with these values of $\mu_0$ are shown in
 Fig.~\ref{farir}, compared to the same NLO, NNLO and
 resummed cross sections as presented before. 
The ratios of the infrared-regulated resummed 
cross sections to LO show a smoother increase than the 
``purely minimally'' resummed ones. 
The difference is particularly marked at the lower center of mass
 energy in Fig.~\ref{farir} (left), with only a modest enhancement
 over NNLO remaining. We interpret these results to indicate a
strong sensitivity to nonperturbative dynamics at the lower energies,
and much less at the higher. 

\section{SPIN ASYMMETRY $A_{TT}$}
To perform numerical studies of the asymmetry 
$A_{TT}$ we need to make a model for the
 transversity densities in the valence region. 
Here, guidance is provided by the Soffer inequality \cite{ref:soffer}
\begin{equation}
\label{eq:sofferineq}
2\left|\delta q(x,Q^2)\right| \leq q(x,Q^2) + \Delta q(x,Q^2) \; ,
\end{equation}
 which gives an upper bound for each $\delta q$. 
Following~\cite{attnlo} we utilize this inequality by saturating 
 the bound at some low input scale $Q_0\simeq 0.6\,\mathrm{GeV}$ using 
 the NLO GRV~\cite{grv} and GRSV (``standard scenario'')~\cite{grsv}
 densities $q(x,Q_0^2)$ and $\Delta q(x,Q_0^2)$, respectively.
For $Q>Q_0$ the transversity densities $\delta q(x,Q^2)$ are then
 obtained using the NLO evolution equations~\cite{wv}. 

\begin{figure}[h]
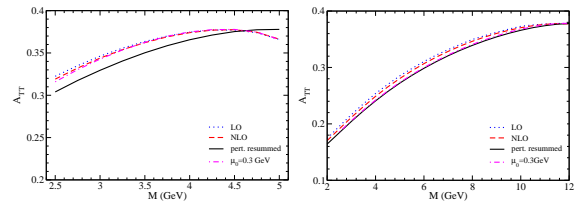

\vspace{-10pt}
\begin{center}
\epsfig{figure=asym-s30-non.eps,width=.23\textwidth}
\epsfig{figure=asym-s210-non.eps,width=.23\textwidth}
\vspace{-40pt}
\caption{Spin asymmetry $A_{TT}(\phi=0)$ at LO, NLO and for the
 NLL-resummed case at $S=30$ GeV$^{2}$ (left) and $S=210$
 GeV$^{2}$ (right).
}\label{asym}
\end{center}
\vspace{-30pt}
\end{figure}

Figure~\ref{asym} shows that $A_{TT}$ is very robust 
 under the QCD corrections, including resummation with and without
a cutoff. This is expected to some extent because the emission of soft-gluons 
does not change the spin of the parent parton. 
\section{NNLL RESUMMATION}
Thanks to the recent calculation of the three-loop splitting functions by
Moch, Vermaseren and Vogt~\cite{3loop}, we can now perform the threshold
resummation for the Drell-Yan process to NNLL accuracy. This 
leads to a new term in the exponent in Eq.~(\ref{lndeltams}):
\begin{eqnarray}
\label{nnllres}
&&\ln \delta \omega_{q\bar{q}}^{{\rm res},N} (r,\as(\mu))
= C_q (r,\as(\mu)) \nn\\ 
&&\hspace{55pt}+ 2 \ln \bar{N} \;h^{(1)}(\lambda) +
2h^{(2)}(\lambda,r)\nn\\
 &&\hspace{55pt}+ 2\alpha_s(\mu)h^{(3)}(\lambda,r)\ ,
\end{eqnarray}
where $h^{(3)}$
includes the new $A^{(3)}_q$ and $D_{DY}^{(2)}$ 
coefficients~\cite{vogt,catani2}, 
and where an additional $C_q^{(2)}$ term is included in
the coefficient function, which may be extracted~\cite{mv1} from the
known~\cite{vN} NNLO results for the Drell-Yan process.
\begin{figure}[t]
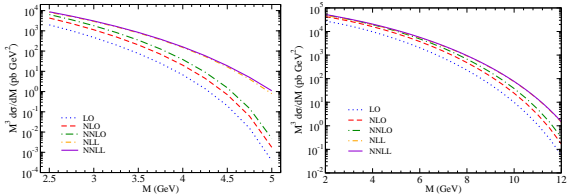

\begin{center}
 \epsfig{figure=s30-nnll-r.eps,width=.23\textwidth}
 \epsfig{figure=s210-nnll-r.eps,width=.23\textwidth}
\vspace{-40pt}
\caption{Unpolarized cross section $M^3 d\sigma/dM$ at $S=30$
 GeV$^{2}$ (left) and $S=210$ GeV$^2$ (right) at LO, NLO, NNLO, and
 NLL-, NNLL-resummed, as function of lepton pair invariant mass $M$.
}\label{unp}
\end{center}
\vspace{-30pt}
\end{figure}
The effects of NNLL resummation on the unpolarized cross section 
are displayed in Fig.~\ref{unp}. One finds that the resummed cross 
section has a fast convergence, even at the lower energy. 

\section{SUMMARY}
We have studied the perturbative QCD corrections to Drell-Yan
 dilepton production in transversely polarized $\bar{p}p$ collisions
 for the kinematical regime of proposed experiments at GSI-FAIR.
We find that the $K$-factor for the available fixed-order corrections, 
and for the all-order NLL soft-gluon resummation, can be very large. In 
contrast, the spin asymmetry is quite stable. We have highlighted
the importance of rather small momentum scales in the resummed exponent
at the lower energies. We have also examined the resummation to NNLL 
and found it to give a rather modest correction. 

\section*{Acknowledgments}
The work of G.S.\ was supported in part
by the National Science Foundation, grants PHY-0098527, PHY-0354776,
and PHY-0354822. W.V.\ is grateful to RIKEN, BNL
and the U.S.\ Department of Energy (contract number DE-AC02-98CH10886) for
providing the facilities essential for the completion of his work.
The work of H.Y.\ was supported in part by a Research Fellowship of
the Japan Society for the Promotion of Science.

\end{document}